# Transport studies of epi-Al/InAs 2DEG systems for required building-blocks in topological superconductor networks

Joon Sue Lee[1], Borzoyeh Shojaei[1,2], Mihir Pendharkar[3], Anthony P. McFadden[3], Younghyun Kim[4], Henri J. Suominen[5], Morten Kjaergaard[5], Fabrizio Nichele[5], Charles M. Marcus[5], Chris J. Palmstrøm[1,2,3]

[1]California NanoSystems Institute, University of California, Santa Barbara, CA 93106
[2]Materials Department, University of California, Santa Barbara, CA 93106
[3]Department of Electrical and computer Engineering, University of California, Santa Barbara, CA 93106
[4]Physics Department, University of California, Santa Barbara, CA 93106
[5]Center for Quantum Devices and Station Q Copenhagen, Niels Bohr Institute, University of Copenhagen, Universitetsparken 5, 2100 Copenhagen, Denmark

**Abstract:**

One-dimensional (1D) electronic transport and induced superconductivity in semiconductor nano-structures are crucial ingredients to realize topological superconductivity. Our approach for topological superconductivity employs a two-dimensional electron gas (2DEG) formed by an InAs quantum well, cleanly interfaced with a superconductor (epitaxial Al). This epi-Al/InAs quantum well heterostructure is advantageous for fabricating large-scale nano-structures consisting of multiple Majorana zero modes. Here, we demonstrate building-block transport studies using a high-quality epi-Al/InAs 2DEG heterostructure, which could be put together to realize the proposed 1D nanowire-based nano-structures and 2DEG-based networks that could host multiple Majorana zero modes: 1D transport using 1) quantum point contacts and 2) gate-defined quasi-1D channels in the InAs 2DEG as well as induced superconductivity in 3) a ballistic Al-InAs 2DEG-Al Josephson junction. From 1D transport, systematic evolution of conductance plateaus in half-integer conductance quanta are observed as a result of strong spin-orbit coupling in the InAs 2DEG. Large $I_c R_n$, a product of critical current and normal state resistance from the Josephson junction, indicates that the interface between the epitaxial Al and the InAs 2DEG is highly transparent. Our results of electronic transport studies based on the 2D approach suggest that the epitaxial superconductor/2D semiconductor system is suitable for realizing large-scale nano-structures for quantum computing applications.

I. INTRODUCTION

Studies of Majorana zero modes, the midgap state localized at the ends of a one-dimensional (1D) topological superconductor [1–3], have gained much attention due to potential applications for fault-tolerant quantum computation [4–6]. Soon after the theoretical proposals of the Majorana zero modes, evidence has been experimentally demonstrated using InSb and InAs nanowires [7–15]. Based on the superconductor/1D nanowire systems, designs for topological quantum computing processes composed of multiple Majorana zero modes, have been proposed [16–19]. One obvious question in realizing the proposed designs is how to make the complex nano-structures by aligning individual 1D nanowires. One promising alternative regarding this question would be to fabricate the nano-structures from s-wave superconductor/two-dimensional (2D) semiconductor systems. In addition to the 1D nanowire-based proposals, recent theoretical works have proposed realization of Majorana zero modes in 2D semiconductor systems using planar Josephson junctions [20] and strips of a proximity-coupled 2D electron gas (2DEG) [21]. Designs of networks of multiple Majorana zero modes for topological quantum computation were also suggested in the latter proposal [21]. The basic components of the proposed superconductor/1D nanowire nano-structures or superconductor/2DEG networks are: a) narrow semiconductor channels of 1D nanowires or narrow strips of 2DEG with a strong spin-orbit coupling, b) transparent interface between s-wave superconductor and semiconductor, c) proximity-induced superconductivity in the semiconductor with an external magnetic field for topological superconducting states, d) a narrow gate on nanowire or a quantum point contact (QPC) in 2DEG for tunneling that controls transfer of electrons in and out of the Majorana zero modes, and e) semiconductor quantum dots for coupling of the Majorana zero modes in 1D nanowire-based proposals.

A recent report on a hybrid superconductor/2D semiconductor heterostructure (epitaxial Al/InAs 2DEG) demonstrated a strong coupling between Al and InAs 2DEG, which is necessary to realize topological superconductivity [22]. In this material system, induced superconductivity in the InAs 2DEG has been extensively studied: a doubling of the lowest quantized plateau and hard superconducting gap in the proximitized 2DEG [23], as well as near-unity transmission at the interface between epitaxial Al and InAs 2DEG by measuring multiple Andreev reflections in a Josephson junction [24]. Most recently, Majorana zero modes have been demonstrated in the epi-Al/InAs 2DEG system [25].

In this work, we prepare an epi-Al/InAs 2DEG heterostructure with improved InAs 2DEG properties by a few times higher electron mobility and lower electron density comparing to those of the epi-Al/InAs 2DEGs used in the recent reports [22–25]. The *in-situ* epitaxial Al layer makes a transparent interface for a strong proximity-induced superconductivity in the InAs 2DEG. We study individual building-block transport components of the 1D nanowire-based nano-structures and 2DEG-based networks using the high-quality epi-Al/InAs 2DEG system. Narrow semiconductor channels with various lengths are fabricated using top gates after removing the Al layer: 1) A short channel forms a QPC where electrons move ballistically, and the QPCs are used to study conductance quantization near the tunneling regime. 2) Narrow 1D channels whose length is similar to the mean free path of the InAs 2DEG or longer, are defined by split gates on the 2DEG to study the nanowire-like 1D transport. Clear quantization of conductance in integer and half-integer units of the conductance quantum is observed in both QPCs and gate-defined qausi-1D channels. We also investigate 3) the induced superconductivity in this epi-Al/InAs 2DEG system using a ballistic Josephson junction. A highly transparent interface is evidenced by a large value of the product of the critical current and the normal resistance, $I_c R_n$. The gate-tunable supercurrent induced in the narrow InAs 2DEG channel persists up to ~800 mT in an in-plane magnetic field.

## II. SAMPLE AND DEVICE PREPARATIONS

The InAs 2DEG was grown on a semi-insulating InP (001) substrate, which is 3.4% lattice-mismatched to InAs, by molecular beam epitaxy (MBE). Initially, the metamorphic, graded buffers of $In_xAl_{1-x}As$ was grown in 18 steps from x=0.52 to x=0.81 at a substrate temperature of ~330˚C. Quantum well (QW) layers of $In_{0.75}Ga_{0.25}As$ (4 nm)/InAs (7 nm)/$In_{0.75}Ga_{0.25}As$ (10 nm) were grown at 460˚C on the top-most buffer layer of 106-nm-thick $In_{0.81}Al_{0.19}As$, grown at 480˚C. A Si-delta-doping layer, of a sheet density of $2 \times 10^{12}$ cm$^{-2}$, was placed 6 nm below the top surface of the buffers. After the QW growth, the wafer was cooled down below room temperature in the growth chamber with all cells were at idle temperatures and with the wafer facing down to a liquid nitrogen cryo-panel. Finally, an epitaxial Al layer of 10 nm was grown on the QW heterostructure [22]. Electronic properties of the InAs 2DEG were characterized using a Hall bar geometry with the Al layer removed: the electron carrier density $n_{2D}$ is $8.7 \times 10^{11}$ cm$^{-2}$ with the electron mobility of 52,400 cm$^2$/V·s at 2 K. All the devices for 1D transport and superconducting transport studies described in this work are fabricated from this epi-Al/InAs QW wafer.

For 1D transport studies, QPCs and gate-defined quasi-1D channels were fabricated on 10-μm-wide InAs channels. First, the wide epi-Al/InAs QW channels were defined by electron-beam lithography and wet etching. Then the epitaxial Al layer on the channels was selectively removed by wet etching (using Transene D etchant at 50˚C), and the whole wafer was covered by 40-nm-thick $Al_2O_3$ as a gate dielectric using atomic layer deposition (ALD). Gate metals of Ti/Au (10 nm/90 nm) were deposited by electron-beam evaporation to form the electron-beam-patterned QPCs and gate-defined quasi-1D channels.

For superconducting transport studies, a superconductor/non-superconductor/superconductor (S-N-S) junction of Al-InAs 2DEG-Al was fabricated with a top gate. Similarly to previously described fabrication processes for QPCs and gate-defined quasi-1D channels, a 1.5-μm-wide epi-Al/InAs 2DEG channel was initially fabricated, followed by a selective wet-etch of an epitaxial Al strip to form an S-N-S junction with the two S contacts separated by bare, 40-nm-long InAs 2DEG region. In order to study gate-voltage dependence of the induced superconductivity in the bare semiconductor region, a 30-nm-thick $HfO_2$ gate dielectric was blanket-deposited by ALD, and Ti/Au (5 nm/25 nm) gate metals were deposited by electron-beam evaporation after an electron-beam patterning of a gate.

## III. EXPERIMENTAL RESULTS

### A. Quantized conductance through QPCs

Conductance through QPCs on the near-surface InAs 2DEG is quantized in integer units and half-integer units of the conductance quantum ($G_0 = 2e^2/h$). Figure 1(a) shows a scanning electron microscopy (SEM) image of a QPC device (Device1) on the InAs QW. An AC excitation current of 10 nA is applied through the InAs 2DEG channel while measuring differential voltage across the QPC in a magnetic field perpendicular to the wafer plane [Fig. 1(b)]. At zero magnetic field, conductance with symmetric split-gate voltages ($V_{sg} \equiv V_{sg1} = V_{sg2}$) shows plateaus of $1G_0$ and $2G_0$ [black curve in Fig. 1(c)]. By applying a perpendicular magnetic field, energy spacing between subbands increases, and the conductance plateaus widen, referred to the magneto-electric depopulation [26,27]. At sufficiently strong magnetic fields, additional half-integer values of the conductance quantum appear as the spin degeneracy of the subbands is lifted due to Zeeman spin splitting. Conductance curves at 6 T (blue) and 8 T (magenta) in Fig. 1(c) clearly show the widening of the conductance plateaus and the emerging half-integer plateaus of $0.5G_0$, $1.5G_0$, and $2.5G_0$. We note that in addition to integer and half-integer conductance quanta, a shoulder near $0.7G_0$ was observed at zero magnetic field in Fig. 1(c), which may be the 0.7 anomaly [28]. The conductance steps are rounded at higher temperatures because electrons in a subband partially occupy adjacent subbands by excitations due to the thermal energy $k_B T$ [Fig. 1(d)].

We also observe an evolution of half-integer conductance quantum as the lateral potential confinement in the narrow constriction of a QPC becomes highly asymmetric. In a second QPC device (Device2) with a similar geometry, the gate voltage sweep was applied for symmetric ($V_{sg1} = V_{sg2}$) and asymmetric ($V_{sg1} \neq V_{sg2}$) lateral potential confinements. As shown in Fig. 1(e), the conductance curves as a function of gate voltage with a constant voltage difference $\Delta V_{sg} = V_{sg1} - V_{sg2}$, were measured at 2 K in a perpendicular magnetic field of 9 T. For a symmetric gate voltage sweep ($\Delta V_{sg} = 0$ V, black curve), integer conductance quanta of $1G_0$ and $2G_0$ as well as half-integer conductance quantum of $1.5G_0$ due to Zeeman splitting, are clearly seen. A slight kink of $0.5G_0$ at $\Delta V_{sg} = 0$ V becomes more pronounced as the lateral potential confinement becomes more asymmetric (up to $\Delta V_{sg} = 7$ V). No half-integer conductance quantum either by Zeeman splitting or by applying asymmetric lateral potential is observed at a magnetic field below 5 T (not shown in the figure). One possible interpretation of the $0.5G_0$ by the asymmetric lateral potential is quantization as a result of spontaneous spin polarization induced by lateral spin-orbit coupling [29,30]. However, the $0.5G_0$ is not always due to the spin polarization. A theoretical study showed that $0.5G_0$ may be due to multichannel interference even in the absence of spin-orbit interaction [31]. The conductance plateaus are also sensitive to surface roughness and impurity scattering [32]. In our case, a magnetic field of 9 T is applied, which already breaks the spin degeneracy. Applying asymmetric gate voltages modifies the lateral potential landscape and changes channels for electron conduction. The visibility of the $0.5G_0$, split by Zeeman effect, could be enhanced by changing the lateral potential landscape. The surface scattering, which is a dominant scattering mechanism in the near-surafce InAs 2DEG [22], could also play a role for the evolution of the $0.5G_0$. Further studies would be necessary using deep InAs 2DEG systems of significantly less scattering centers to verify the effect of the surface scattering and the multichannel interference.

### B. Gate-defined quasi-1D channels

We study electrical transport through quasi-1D channels defined by electrostatic gating. Two channels are prepared with a similar width of ~110 nm and different lengths of 1.13 μm for Device3 [Fig. 2(a) inset] and 2.16 μm for Device4. Conductance through a gate-defined channel at zero magnetic field is not quantized, as shown in Fig. 2(a). The long channels, which are 5 or 10 times longer than those of the QPCs, lead to more scattering of electrons, smearing the conductance quantization at zero magnetic field. We note that the mean free path of the 2DEG is 806 nm, which is shorter than the length of the channels. In contrast, with a strong magnetic field, conductance quantization appears at integer values as well as at half-integer values of the conductance quantum [Fig. 2(b)]. Application of magnetic field increases the spacing between spin-degenerated subbands and Zeeman-spin-splits each of them.

The scattering of electrons in the gate-defined channels lead to weak localization (WL) corrections to the conductance [33] due to random potentials. Temperature dependence of magneto-transport of the gate-defined quasi-1D channels shows resistance peaks at zero magnetic field from the WL corrections [Fig. 2(c)]. The WL near zero magnetic field persists in the measurement range of 2 – 24 K while the conductance fluctuations [34] at higher fields are suppressed significantly as temperature increases. We study the WL effect more closely by tuning the gate voltage ($V_{sg}$) at 16 K where the magnitude of the conductance fluctuations is much lower than that of the WL corrections. Application of the gate voltage determines the number of the conduction channels. In Fig. 2(d), the conductance curves in the range of $V_{sg}$ = -5.32 V ~ -5.22 V are in quasi-1D limit of a few conduction channels while the conductance curves of $V_{sg}$ = -5.0 V and -4.8 V are in 2D limit of multiple conduction channels (> 10 conduction channels). The WL corrections are fit by an equation [35]:

$$\mathrm{dG}\,(B) = \frac{2e^2}{h}\frac{1}{L}\left(L_\phi - \left(L_\phi^{-2} + L_B^{-2}\right)^{-\frac{1}{2}}\right) \text{ with } L_B = l_m\sqrt{\left(\frac{C_1 l_m^2 l_e}{w^3} + \frac{C_2 l_e^2}{w^2}\right)}, \quad (1)$$

where $l_m = \sqrt{\hbar/eB}$, $C_1 = 2\pi$, $C_2 = 1.5$, $l_e$ is the mean free path, $w$ is the width, and $L_\phi$ is the dephasing length. Here we fixed the mean-free path and diffusion constant for the quasi-1D channels to be the same value as for the 2DEG, and fit the data with the only one free parameter, $L_\phi$. The resulting coherence length is shorter in the quasi-1D limit than that in the 2D limit [Fig. 2(e)]. Therefore, we can interpret the result as that depleting the channel leads to stronger inelastic process, such as electron-electron interaction induced dephasing [36].

C. Induced superconductivity in a ballistic Josephson junction with a top gate

An S-N-S Josephson junction of Al-InAs 2DEG-Al is prepared to investigate the transparency of the interface between the epitaxial Al layer and the InAs 2DEG by the proximity-induced superconductivity in the InAs 2DEG. A clean, transparent interface between a superconductor (S) and a normal non-superconductor (N) results in a strong proximity effect. The Al-etched InAs 2DEG junction is 1.5 μm wide and 40 nm long and covered by a top gate, as illustrated in Fig. 3(a). Note that the length of the junction is an order of magnitude shorter than the mean free path of the InAs 2DEG, so that the 40-nm-long junction is in a ballistic regime. The critical temperature of the Al film is ~1.5 K with the corresponding superconducting gap, $\Delta = 1.75 k_B T_c \cong 228\ \mu eV$. Below the critical temperature, differential resistance and the DC voltage across the S-N-S junction as a function of DC current show a clear indication of the induced superconductivity—supercurrent, electric current without dissipation [Figs. 3(b) and 3(c)].

The supercurrent is tunable by the top gate and magnetic field. As the InAs 2DEG is depleted by applying negative gate voltages, supercurrent is suppressed and eventually disappears at $V_g$ = -2.1 V [Fig. 4(a)]. In an in-plane magnetic field, the supercurrent persists up to ~800 mT [Fig. 4(b)]. In previous reports where Al was used as a superconductor that proximitizes InAs or InSb nanowires, the zero-energy anomalies from Majorana bound states have been demonstrated with an external magnetic field in a range of 35 - 2000 mT [7,8,14,15]. This confirms that the induced superconductivity in the epi-Al/InAs 2DEG system is usable for experiments investigating Majorana zero-energy modes. Given the fact that Majorana zero modes have been demonstrated [25] in an epi-Al/InAs 2DEG system of inferior transport properties to the system studied in this paper, studies of topological superconductivity with less disorder is expected using this epi-Al/InAs 2DEG system.

## IV. DISCUSSIONS

One of the advantages of 2DEGs over 1D nanowires for Majorana devices is a potential for higher electron mobility. In the case of deep InAs quantum wells buried deeply from the surface, the electron mobility of near or higher than $10^6$ cm$^2$/V·s has recently been reported [37,38]. In contrast, the epi-Al/InAs quantum well heterostructures, where the top barrier on InAs layer is engineered to be thin to sustain the wavefunction overlap between the Al layer and the InAs layer, have relatively low electron mobility (~$10^4$ cm$^2$/V·s) mainly due to the surface scattering [22]. In this work, we carry out transport studies on an InAs quantum well with an improved electron mobility approximately by a factor of three with a lower electron density. The improved transport properties result in clean conductance quantization in narrow InAs 2DEG channels as well as strong induced superconductivity in a Josephson junction.

Studies of QPCs and fabricated 1D channels in near-surface InAs quantum wells have been lacking while those in deep InAs quantum wells have been demonstrated [29,39,40]. Our results of clean conductance quantization, despite of the relatively lower quality in the near-surface InAs 2DEGs, show great promise as building blocks in networks of topological superconducting materials. Systematic evolution of $0.5G_0$, $1.5G_0$, and $2.5G_0$ was observed due to Zeeman splitting by applying perpendicular magnetic field. Also, we observe the enhancement of $0.5G_0$ as the lateral potential confinement in a QPC becomes more asymmetric. Strong spin-orbit coupling in InAs is the motivation of using InAs 2DEG for realizing topological superconductivity. Further improvement of the InAs 2DEG quality may allow one to observe the helical gap [41] where spin-orbit interaction with effective magnetic field significantly modifies the band structure of a narrow 1D channel.

A self-assembled-nanowire-like 1D channel can be realized in a 2DEG by two different ways: 1) a gate-defined 1D channel by depleting 2DEG area underneath the two top, split gates, and 2) an etch-defined 1D channel by etching the 2DEG to form a narrow strip. For electronic transport studies, gate-defined 1D channel is more viable since the etch-defined 1D channel has more effect of disorder from potential barriers of the etched side walls [42]. We focus only on the gate-defined 1D channel in this study. The coherence length of ~400 nm, obtained from analyzing WL in the 1D limit, is comparable to that of self-assembled InAs nanowires [43]. We note that in this paper we do not investigate the transport through a quantum dot, which is one of the building-blocks for 1D-nanowire based networks of topological superconductors. A combination of what we have demonstrated, a gate-defined 1D channel and QPCs for tunneling, can form a quantum dot.

In the ballistic Al-InAs 2DEG-Al Josephson junction, supercurrent breaks down with a current higher than critical current, $I_c = 1.34$ $\mu A$ at 30 mK. A product of the critical current $I_c$ and the normal state resistance $R_n$ is proportional to the gap size, $I_c R_n = \alpha \Delta/e$, and generally used as a junction quality factor. It is given by $I_c R_n = (\pi/2)(\Delta/e) \tanh(\Delta/2kT)$ near $T_c$ for an S-N-S junction of a metallic weak link [44,45]. Using theories that are valid all the way to $T = 0$ K, the numerical factor $\alpha$ approaches $1.32(\pi/2)$ and $2(\pi/2)$ at $T = 0$ K for the dirty and clean limits [46,47]. From our measurements, $I_c R_n = 1.34$ $\mu A \times 225$ $\Omega \cong 302$ $\mu eV$ gives $\alpha = 1.32$, which is 64% of the dirty limit and 50% of the clean limit. The excess current $I_{exc}$ is an additional current to the normal current ($I_{exc} = I - V/R_n$) as a result of the Andreev reflection at S-N interfaces [48], and it linearly scales with the gap size, $I_{exc} R_n = \alpha' \Delta/e$. For ideal, transparent S-N interfaces, the numerical factor becomes $\alpha' = 8/3$ in the case of ballistic junctions [48] while $\alpha' = (\pi^2/4 - 1)$ in the case of diffusive junctions [49]. We extract the excess current, $I_{exc} = 1.27$ $\mu A$, by extrapolating a linear fit of the I-V curve in the normal resistance regime ($V \gg \Delta/e$) to $V = 0$ mV, as shown in Fig. 3(c). The resulting $I_{exc} R_n$ equals to 287 $\mu eV$ with $\alpha' = 1.26$, which is 86% of the diffusive junction limit and 47% of the ballistic junction limit. Both $I_c R_n$ and $I_{exc} R_n$ are approximately ~50% to the theoretical calculations for clean and ballistic junction limits. These numbers are highest to date in Al-semiconductor based Josephson junctions, to the best of our knowledge. The high $I_c R_n$ and $I_{exc} R_n$ indicate a transparent interface between the epitaxial Al and the high-quality InAs 2DEG.

## V. CONCLUSIONS

Using an MBE-grown high-quality epi-Al/InAs 2DEG heterostructure, we have demonstrated electronic transport studies that could form basic components of theoretically proposed complex nano-structures consisting of multiple Majorana zero modes. For tunneling regime transport, we investigate QPCs that could completely deplete the narrow constriction in InAs 2DEG and tune the conductance in integer and half-integer conductance quanta by applying symmetric/asymmetric gate voltage and magnetic field. Transport through narrow 1D channels was demonstrated by gate-defined quasi-1D channels of length comparable with or longer than the coherence length of the 2DEG. Quantized conductance with a magnetic field as well as systematic change in the coherence length by applying gate voltage was observed. Induced superconductivity was investigated by a ballistic Al/InAs 2DEG/Al Josephson junction with a top gate. High $I_c R_n$ and $I_{exc} R_n$ indicate a clean, transparent interface. Supercurrent induced in the narrow InAs 2DEG channel is gate-tunable and persists up to ~800 mT in-plane magnetic field. Our transport studies suggest that this epitaxial superconductor/2D semiconductor system could work as a promising platform to large-scale nano-structures utilizing multiple Majorana zero modes for quantum computing applications.

ACKNOWLEDGEMENTS

This work was supported by Microsoft Research and California NanoSystems Institute in Santa Barbara. Part of this work was carried out in UCSB Nanofabrication facility.

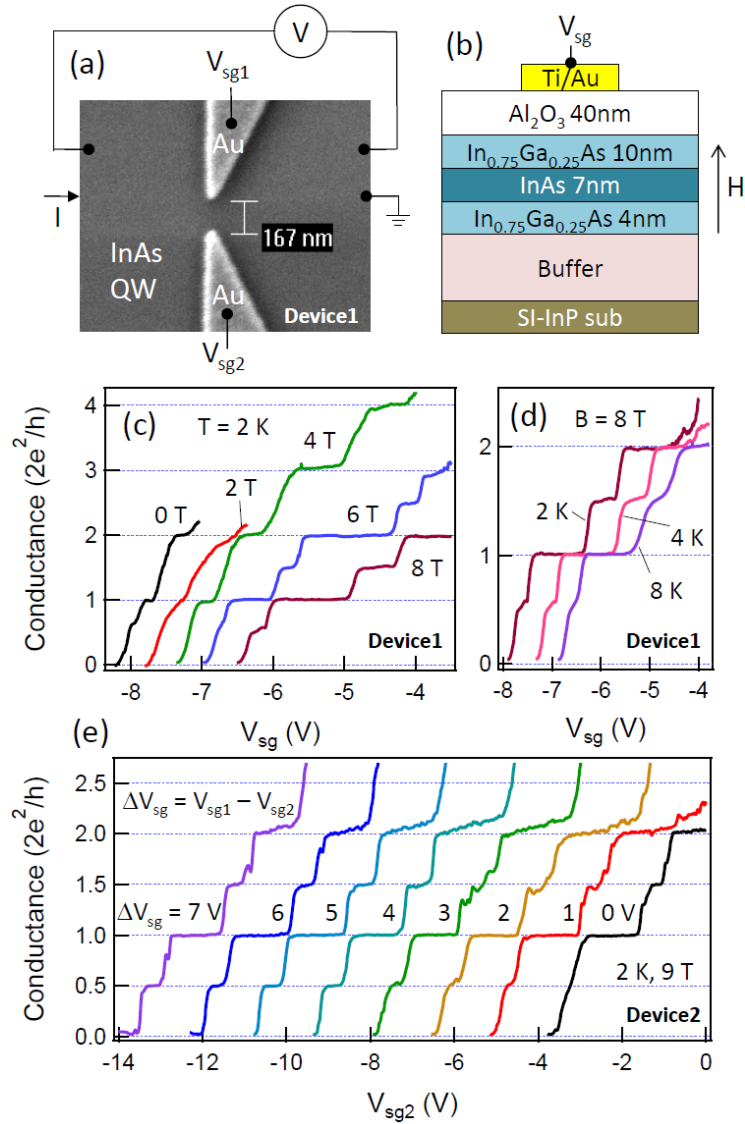

Figure 1. (a) An SEM image and a measurement setup of a QPC device (Device1) on InAs QW. (b) A cross-sectional schematic of the InAs QW heterostructure with ALD $Al_2O_3$ dielectric and Ti/Au gate metal for top-gating. (c) Magnetic field dependence of conductance through a QPC (Device1) as a function of $V_{sg}$ shows conductance plateaus with various perpendicular magnetic fields of 0 T (black), 2 T (red), 4 T (green), 6 T (blue), and 8 T (magenta). Successive curves offset by 0.35 V, with no offset on the curve corresponding to 0 T. (d) Temperature dependence of conductance plateaus at 2 K (magenta), 4 K (pink), and 8 K (purple). Successive curves offset by 0.5 V, with no offset on the curve corresponding to 2 K. (e) Asymmetric lateral potential confinement induces an evolution of $0.5G_0$ plateau. Successive curves offset by 0.7 V, with no offset on the curve corresponding to $\Delta V_{sg} = 7$ V.

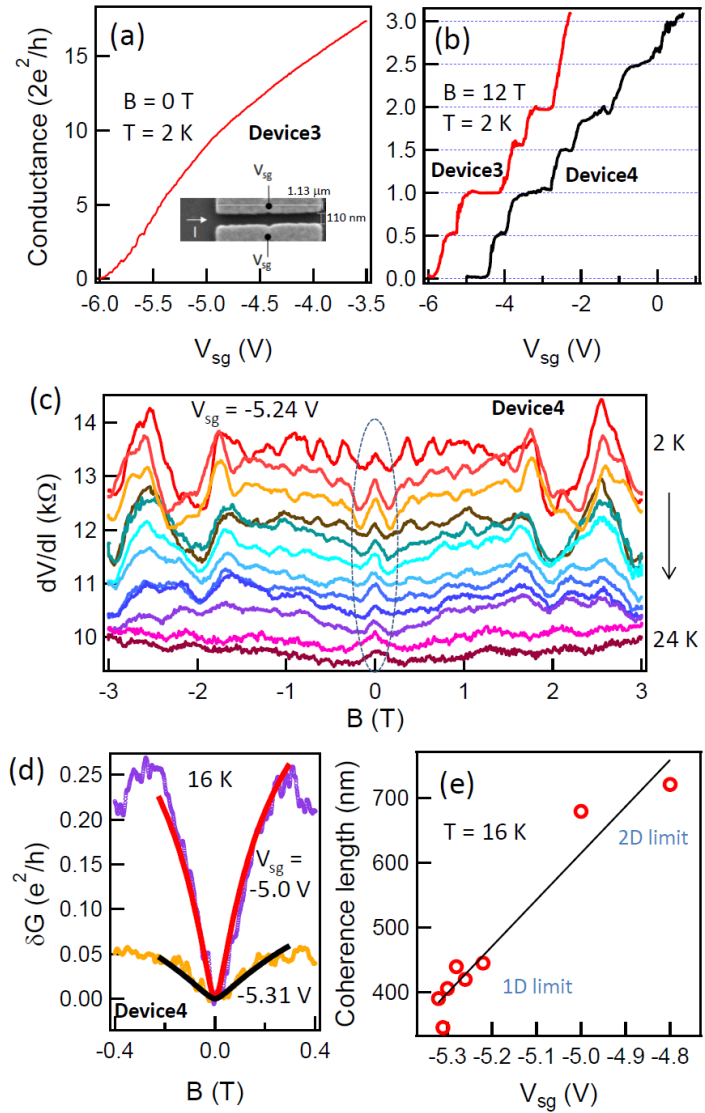

Figure 2. (a) Conductance as a function of $V_{sg}$ in a gate-defined channel (Device3) with zero magnetic field at 2 K. Inset shows an SEM image of Device3. (b) With a high magnetic field (12 T), conductance through the gate-defined channels from Device3 (red) and Device4 (black) reveals conductance quantization. Data from Device4 is offset in $x$-axis for clarity. (c) Magneto-resistance across the gate-defined channel (Device4) at every 2 K from 2 K to 24 K. Offset for clarity. Dotted circle highlights WL near zero magnetic field. WL corrections at 16 K (d) Magneto-conductance curves at $V_{sg}$= -5 V (purple) and -5.31 (yellow). Red and black solid lines are fitted lines by Eq. (1). (e) Resulting coherence length from WL fitting at various $V_{sg}$ in 1D limit and in 2D limit. Black line is a linear fit.

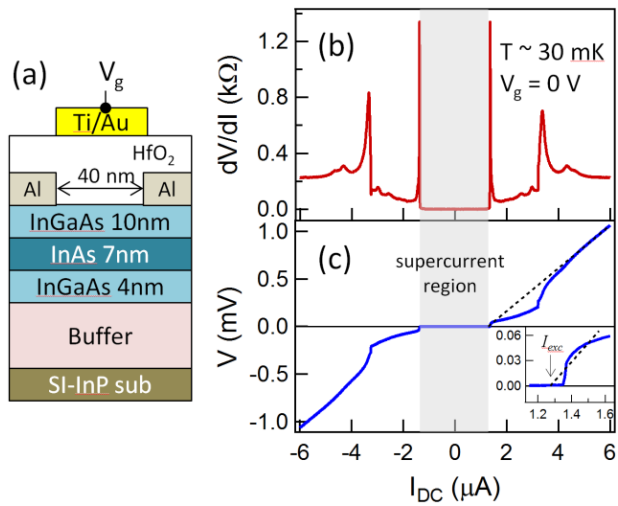

Figure 3. (a) Cross-sectional schematic of an Al-InAs 2DEG-Al junction with a top gate. (b) Differential resistance and (c) DC voltage across the junction as a function of DC current with zero gate voltage at ~30 mK. Grey region represents a DC current range (±1.35 µA) where supercurrent exists. Black dotted line is a linear fit to the large voltage region ($V \gg \Delta/e$) of normal resistance.

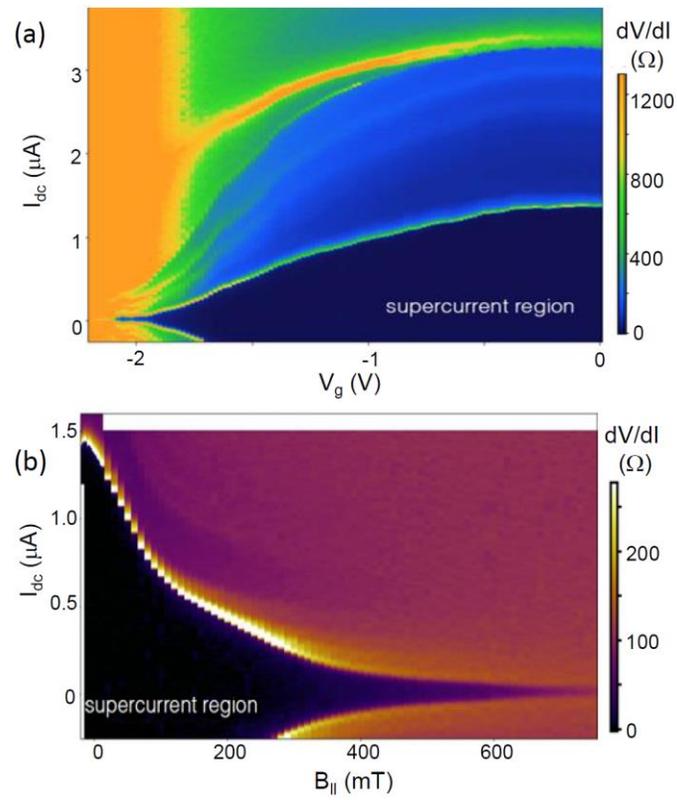

Figure 4. (a) Differential resistance vs DC current $I_{DC}$, measured as a function of top gate voltage. (b) Differential resistance vs DC current $I_{DC}$, measured as a function of in-plane magnetic field. All measurements were done at ~30 mK.